\documentclass[twocolumn,showpacs,aps,prl,superscriptaddress]{revtex4}

\usepackage{graphicx}
\usepackage{dcolumn}
\usepackage{epsfig}
\usepackage{psfrag}
\usepackage{amsmath}

\newcommand{\BaBarYear}    {08}
\newcommand{\BaBarNumber}  {015}

\newcommand{\SLACPubNumber} {13231}
\newcommand{\LANLNumber} {0805.1217 [hep-ex]}

 \newcommand{\BaBarType}      {PUB}  

\input pubboard/babarsym   

\RequirePackage{xspace}

\newcommand{\pvec}{{\bf p}}

\newcommand{\acp}{\ensuremath{\calA_{ch}}}

\newcommand{\calB}{\ensuremath{{\cal B}}}

\newcommand{\DE}{\ensuremath{\Delta E}}

\newcommand{\xf}{\ensuremath{{\cal F}}}

\newcommand{\thetaT}{\ensuremath{\theta_{\rm T}}}
\newcommand{\costhr}{\ensuremath{\cos\thetaT}}

\newcommand\etal{{\it et al.}}
\newcommand{\half}{\ensuremath{\frac{1}{2}}}

\newcommand{\bfig}{\begin{figure}[htbpc!]}
\newcommand{\efig}{\end{figure}}
\newcommand\bef{\begin{figure}}
\newcommand\edf{\end{figure}}
\newcommand\dbline{\noalign{\vskip 0.10truecm\hrule}\noalign{\vskip 2pt}\noalign{\hrule\vskip 0.10truecm}}
\providecommand{\tbline}{\noalign{\vskip 0.05truecm\hrule\vskip0.05truecm}}

\newcommand\beq{\begin{equation}}
\newcommand\eeq{\end{equation}}
\newcommand\bear{\begin{array}}
\newcommand\enar{\end{array}}
\newcommand\beqa{\begin{eqnarray}}
\newcommand\eeqa{\end{eqnarray}}
\newcommand\ben{\begin{enumerate}}
\newcommand\een{\end{enumerate}}

\newcommand{\UfourS}{\ensuremath{\Upsilon(4S)}}

\newcommand{\bone}{\ensuremath{b_1}}
\newcommand{\bonep}{\ensuremath{b_1^+}}

\newcommand{\bonez}{\ensuremath{b_1^0}}

\newcommand{\fbppiz}{\ensuremath{\bonep \piz}\xspace}
\newcommand{\bppiz}{\ensuremath{\Bp\ra\fbppiz}\xspace}
\newcommand{\Bbppiz}{\ensuremath{\calB(\bppiz)}\xspace}

\newcommand{\fbzpiz}{\ensuremath{\bonez \piz}\xspace}
\newcommand{\bzpiz}{\ensuremath{\Bz\ra\fbzpiz}\xspace}
\newcommand{\Bbzpiz}{\ensuremath{\calB(\bzpiz)}\xspace}

\newcommand{\fbpKz}{\ensuremath{\bonep \Kz}\xspace}
\newcommand{\bpKz}{\ensuremath{\Bp\ra\fbpKz}\xspace}
\newcommand{\BbpKz}{\ensuremath{\calB(\bpKz)}\xspace}

\newcommand{\fbzKz}{\ensuremath{\bonez \Kz}\xspace}
\newcommand{\bzKz}{\ensuremath{\Bz\ra\fbzKz}\xspace}
\newcommand{\BbzKz}{\ensuremath{\calB(\bzKz)}\xspace}

\newcommand{\rbpKz}{\ensuremath{9.6\pm 1.7\pm 0.9}}
\newcommand{\sbpKz}{\ensuremath{6.3}}
\newcommand{\abpKz}{\ensuremath{-0.03\pm 0.15\pm 0.02}}

\newcommand{\rbzKz}{\ensuremath{5.1\pm 1.8\pm 0.5}}
\newcommand{\sbzKz}{\ensuremath{3.4}}
\newcommand{\ulbzKz}{\ensuremath{7.8}\xspace}

\newcommand{\rbppiz}{\ensuremath{1.8\pm 0.9\pm 0.2}}
\newcommand{\sbppiz}{\ensuremath{1.6}}
\newcommand{\ulbppiz}{\ensuremath{3.3}\xspace}

\newcommand{\rbzpiz}{\ensuremath{0.4\pm 0.8\pm 0.2}}
\newcommand{\sbzpiz}{\ensuremath{0.5}}
\newcommand{\ulbzpiz}{\ensuremath{1.9}\xspace}

\newcommand{\theTitle}{{\boldmath Observation of \bpKz\ and search for
\B-meson decays to \fbzKz\ and $b_1\piz$}} 

\begin{document}


\begin{flushleft}
\babar-\BaBarType-\BaBarYear/\BaBarNumber \\
SLAC-PUB-\SLACPubNumber \\
arXiv:\LANLNumber
\end{flushleft}


\title{\theTitle}

%
\author{B.~Aubert}
\author{M.~Bona}
\author{Y.~Karyotakis}
\author{J.~P.~Lees}
\author{V.~Poireau}
\author{E.~Prencipe}
\author{X.~Prudent}
\author{V.~Tisserand}
\affiliation{Laboratoire de Physique des Particules, IN2P3/CNRS et Universit\'e de Savoie, F-74941 Annecy-Le-Vieux, France }
\author{J.~Garra~Tico}
\author{E.~Grauges}
\affiliation{Universitat de Barcelona, Facultat de Fisica, Departament ECM, E-08028 Barcelona, Spain }
% ...
\collaboration{The \babar\ Collaboration}
\noaffiliation

\date{\today}

\begin{abstract}
We present the results of searches for decays of \B\ mesons to final
states with a \bone\ meson and a neutral pion or kaon.  The data,
collected with the \babar\ detector at the Stanford Linear Accelerator
Center, represent 465 million \BB\ pairs produced in \epem\
annihilation.  The results for the branching fractions
are, in units of $10^{-6}$,
    $\BbpKz=\rbpKz$,
    $\BbzKz=\rbzKz~(<\ulbzKz)$,
    $\Bbppiz=\rbppiz~(<\ulbppiz)$, and
    $\Bbzpiz=\rbzpiz~(<\ulbzpiz)$,
with the assumption that ${\cal B}(\bone\ra\omega\pi)=1$.
We also measure the charge asymmetry
$\acp(\bpKz) = \abpKz$.
The first error quoted is statistical, the second systematic, and the
upper limits in parentheses indicate the 90\%\ confidence level. 
\end{abstract}


\maketitle


Recent searches for decays of \B\ mesons to final states with an
axial-vector meson and a pion or kaon have revealed modes with
branching fractions that are rather large among charmless decays:
$(15-35)\times10^{-6}$ for $\B\ra a_1(\pi, K)$ 
\cite{BaBar_a1pi, BaBar_a1K}, and $(7-11)\times10^{-6}$ for charged pion
and kaon in combination with a \bonez\ or a \bonep\ meson
\cite{BaBar_b1h, conjugate}.
In this paper we present the results of investigations of the
remaining charge states with \bone\ accompanied by a \piz\ or \Kz.  No
previous searches for these modes have been reported.

The mass and width of the \bone\ meson are $1229.5\pm 3.2$ MeV and
$142\pm 9$ MeV, respectively, and the dominant decay is to $\omega\pi$
\cite{PDG2006}.  In the quark model the $b_1$ is the $I^G=1^+$ member of
the $J^{PC}=1^{+-},\ \relax^1\!P_1$ nonet.  The Cabibbo-favored
amplitudes that mediate these decays are those represented by
color-suppressed tree diagrams for the modes with \piz, and ``penguin''
loop diagrams for those with \Kz.  Because the \bone\ meson has even
$G$-parity, only amplitudes in which the \bone\ contains the spectator
quark from the \B\ meson are allowed, apart from isospin-breaking
effects \cite{weinberg}.  Direct \CP\ violation would be indicated by a
non-zero value of the asymmetry $\acp \equiv
(\Gamma^--\Gamma^+)/(\Gamma^-+\Gamma^+)$ in the rates
$\Gamma^\pm(B^\pm\ra F^\pm)$ for charged \B-meson decays to final states
$F^\pm$.

The available theoretical estimates of the branching fractions of \B
mesons to $\bone\pi$ and $\bone K$ come from calculations based on na\"{i}ve
factorization \cite{laporta,calderon}, and on QCD factorization
\cite{ChengYang}.  The latter incorporate light-cone distribution
amplitudes evaluated from QCD sum rules, and predict branching fractions
in quite good agreement with the measurements for $\B\ra\bone\pip$ and
$\B\ra\bone\Kp$ \cite{BaBar_b1h}.  The expected branching fractions from QCD
factorization are about
10$\times10^{-6}$ for \bpKz, and 3$\times10^{-6}$ or less for \bzKz\ and
$\B\ra\bone\piz$ \cite{ChengYang}.

The data for these measurements were collected with the \babar\
detector~\cite{BABARNIM} at the PEP-II asymmetric $e^+e^-$ collider
located at the Stanford Linear Accelerator Center.  An integrated
luminosity of 424 \invfb, corresponding to $(465\pm5)\times 10^6$ \BB\
pairs, was produced by \epem\ annihilation at the $\Upsilon (4S)$
resonance (center-of-mass energy $\sqrt{s}=10.58\ \gev$).
Charged particles from the \epem\ interactions are detected, and their
momenta measured, by a combination of five layers of double-sided
silicon microstrip detectors and a 40-layer drift chamber, both
operating in the 1.5~T magnetic field of a superconducting
solenoid. Photons and electrons are identified with a CsI(Tl)
electromagnetic calorimeter (EMC).  Further charged particle
identification (PID) is provided by the average energy loss ($dE/dx$) in
the tracking devices and by an internally reflecting ring imaging
Cherenkov detector (DIRC) covering the central region.  A detailed
Monte Carlo program (MC) is used to simulate the \B
production and decay sequences, and the detector response
\cite{geant}.

\begin{figure}
\psfrag{FFF}{{$\cal F~~~~$}}
\includegraphics[width=1.0\linewidth]{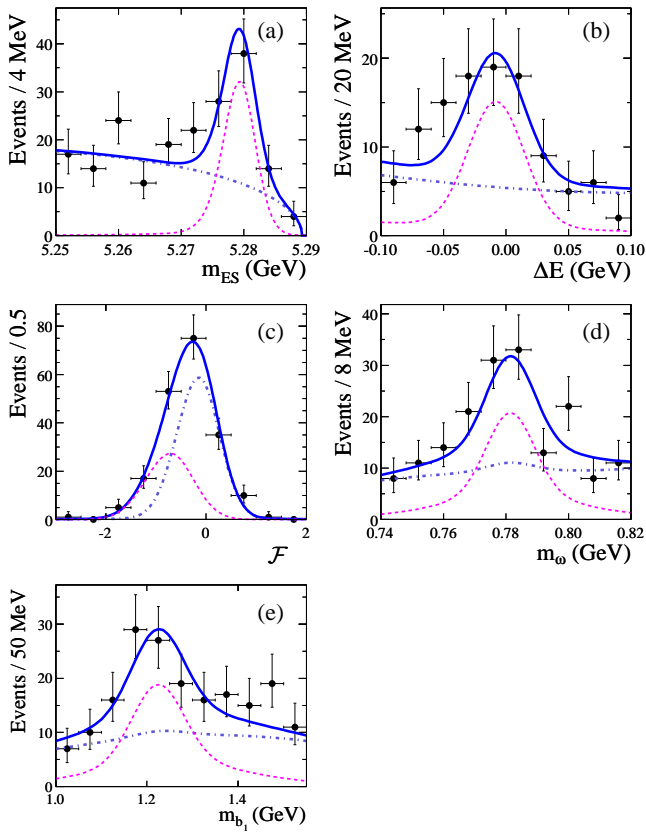}
\caption{\label{fig:proj_b1Ks}
Distributions for signal-enhanced subsets (see text)
of the data projected onto the fit observables for the decay \bpKz;
(a) \mes, (b) \DE, (c) \xf, (d) $m(\pip\pim\piz)$ for the $\omega$
candidate, and (e) $m(\omega\pi)$ for the \bone\ candidate.
The solid lines represent the results of the fits, and the dashed and
dot-dashed lines the signal and background contributions respectively.
}
\end{figure}

The \bone\ candidates are reconstructed through the decay sequence
$b_1\ra\omega\pi$, $\omega\ra\pip\pim\piz$, and $\piz\ra\gaga$.  The
other primary daughter of the \B\ meson is reconstructed as either
$\KS\ra\pip\pim$ or $\piz\ra\gaga$.  For \KS, the invariant mass of the
pion pair is required to lie between 486 and 510 MeV, i.e., within
about 3.5 standard deviations of the nominal \KS\ mass \cite{PDG2006}.
The minimum energy for a \piz-daughter photon is $30\ \mev$ ($50\
\mev$ for a primary \piz), and the minimum energy of a \piz\ is $250\
\mev$.  The invariant mass of the photon pair is required to lie between
120 and 150 MeV, or within about two standard deviations of the
nominal \piz\ mass.  For the \bone\ and $\omega$, whose masses
are treated as observables in the maximum likelihood (ML) fit described
below, we accept a range that includes wider sidebands (see Fig.\
\ref{fig:proj_b1Ks}).  Secondary charged pions in \bone\ and $\omega$
candidates are rejected if classified as protons, kaons, or electrons by
their DIRC, $dE/dx$, and EMC PID signatures.  For a \KS\ candidate we
require a successful fit of the decay vertex with the flight direction 
constrained to the pion pair momentum direction, that yields a flight
length greater than three times its uncertainty.

We reconstruct the \B-meson candidate by combining the four-momenta of
a pair of primary daughter mesons, using a fit that constrains all
particles to 
a common vertex and the \piz\ mass to its nominal value.  From the
kinematics of \UfourS\ decay we determine the 
energy-substituted mass $\mes=\sqrt{\frac{1}{4}s-\pvec_B^2}$ and
energy difference $\DE = E_B-\half\sqrt{s}$, where $(E_B,\pvec_B)$ is
the \B-meson four-momentum vector, and all values are expressed in the
\UfourS\ rest frame.  The resolution in \mes\ is $2.4-2.8\ \mev$ and in
\DE\ is 22--46 MeV, depending on the decay mode.  We require $5.25\
\gev<\mes<5.29\ \gev$ and $|\DE|<100\ \mev$.  

We also impose restrictions on the helicity-frame decay angles of the
\bone\ and $\omega$ mesons.  The helicity frame of a meson is defined as
the rest frame of the meson with $z$ axis along the direction of boost
to that frame from the parent rest frame.  For the decay
$\bone\ra\omega\pi$, $\theta_{\bone}$ is the polar angle of the
daughter pion, and for $\omega\ra3\pi$, $\theta_\omega$ is polar angle
of the normal to the $3\pi$ decay plane.  Since many misreconstructed
candidates accumulate in a corner of the $\cos\theta_{\bone}$ {\it vs} 
$\cos\theta_{\omega}$ plane, we require
$\cos\theta_{\bone}\le\mathrm{min}(1.0,\ 1.1-0.5\times|\cos\theta_{\omega}|)$. 

Backgrounds arise primarily from random combinations of particles in
continuum $\epem\ra\qqbar$ events ($q=u,d,s,c$).  We reduce these with
a requirement on the angle \thetaT\ between the thrust axis
\cite{thrust} of the \B
candidate in the \UfourS\ frame and that of the charged tracks 
and neutral calorimeter clusters in the rest of the event (ROE).  The event is
required to contain at least one charged track not associated with the
\B\ candidate.  The distribution is sharply peaked near $|\costhr|=1$
for \qqbar\ jet pairs, and nearly uniform for \B-meson decays.  The
requirement, which optimizes the expected signal yield relative to its
background-dominated statistical error, is $|\costhr|<0.7$.

The average number of candidates found per event in the selected
sample is in the range 1.3 to 1.6 (1.4 to 1.6 in signal MC), depending
on the final state.  We choose the candidate with the largest
confidence level for the \B-meson geometric fit.  

In the ML fit we discriminate further against \qqbar\ background with a
Fisher discriminant \xf\ that combines five variables: 
the polar angles, with respect to the beam axis in the \UfourS\ rest 
frame, of the $B$ candidate momentum and of the $B$ thrust axis; 
the flavor tagging category; and the zeroth and second 
angular moments $L_{0,2}$ of the energy flow, excluding the $B$
candidate, about the $B$ thrust axis.  The tagging category
\cite{ccbarK0} is the class 
of candidate partially reconstructed from the ROE, designed to determine
whether, in a signal event, it represents a \B\ or \Bb\ meson.
The moments are defined by $ L_j = \sum_i
p_i\times\left|\cos\theta_i\right|^j,$ where $\theta_i$ is the angle
with respect to the $B$ thrust axis of track or neutral cluster $i$,
$p_i$ is its momentum, and the sum excludes the $B$ candidate daughters.
The Fisher variable  provides about one standard deviation of
separation between \B\ decay events and combinatorial background.

We obtain yields for each channel from an extended ML
fit with the input observables \DE, \mes, \xf, and the resonance masses
$m_{\bone}$ and $m_\omega$.   
The selected data sample sizes are given in
Table~\ref{tab:results}.  Besides the signal events these samples
contain \qqbar\ 
(dominant) and \BB\ with $b\ra c$ combinatorial background, and a
fraction of cross feed from other charmless \BB\ modes, which we estimate
from the simulation to be (0.5--1.1)\%.  The last include non-resonant
$\omega\pi\pi$, $\omega K\pi$, and modes that have
final states different from the signal, but with similar
kinematics so that broad peaks near those of the signal appear in some
observables.  We account for these with a separate component in the
probability density function (PDF).

The likelihood function is
\begin{eqnarray}
{\cal L} &=& \exp{\left(\smash[b]{-\sum_j Y_j}\right)}
\prod_i^{N}\sum_j Y_j \times \label{eq:likelihood}\\
&&{\cal P}_j (\mes^i) {\cal P}_j(\xf^i) {\cal P}_j (\DE^i) {\cal P}_j
(m_{\bone}^i) {\cal P}_j (m_{\omega}^i),
\nonumber  
\end{eqnarray}
where $N$ is the number of events in the sample, and for each
component $j$ (signal, combinatorial background, or charmless \BB\
cross feed), $Y_j$ is the yield of events and ${\cal P}_j(x^i)$ the PDF
for observable $x$ in event $i$.  The signal component is further  
separated into two components (with proportions fixed in the fit for each
mode) representing the 
correctly and incorrectly reconstructed candidates in events with true
signal, as determined with MC.  The fraction of misreconstructed
candidates is 32-40\%, depending on the mode.  The factored form of the PDF
indicated in Eq.\ \ref{eq:likelihood}\ is a good approximation,
particularly for the combinatorial \qqbar\ component, since we find
correlations among observables in the data (which are mostly
\qqbar\ background) are generally less than 2\%, with none exceeding
5\%.  The effects of this 
approximation are determined in simulation and included 
in the bias corrections and systematic errors discussed below.

We determine the PDFs for the signal and \BB\ background components
from fits to MC samples.  We calibrate the resolutions in \DE\ and
\mes\ with large data control samples of \B\ decays to charmed final states
of similar topology (e.g.\ $B\ra D(K\pi\pi)\pi$, $B\ra D(K\pi\pi)\rho$).
We develop PDFs for 
the combinatorial background with fits to the data from which the signal
region ($5.27\ \gev<\mes<5.29\ \gev$ and $|\DE|<75\ \mev$) has been
excluded. 

\begin{figure}
\psfrag{FFF}{{$\cal F~~~~$}}
\includegraphics[width=1.0\linewidth]{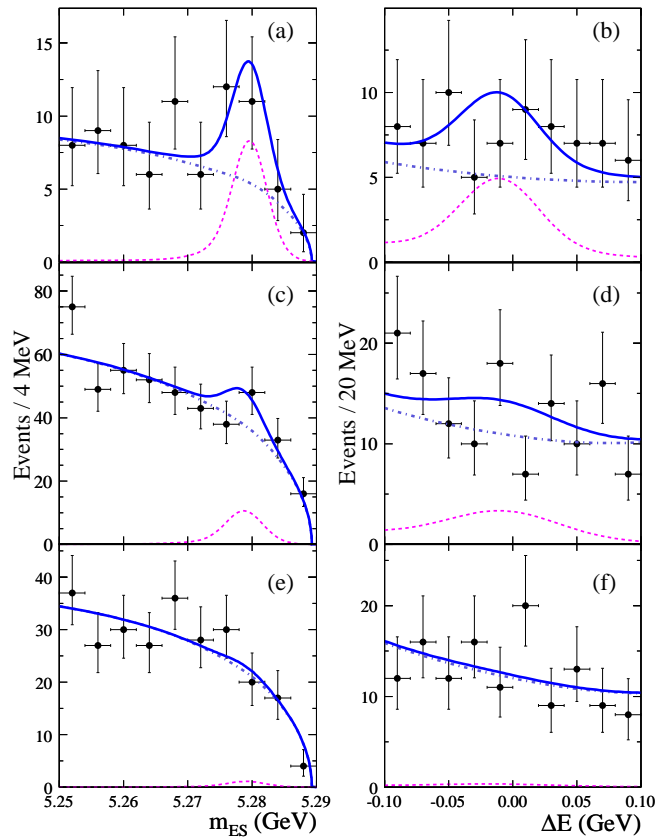}
\caption{\label{fig:proj_all}
Distributions for signal-enhanced subsets (see text)
of the data projected onto \mes\ (a, c, e) and \DE\ (b, d, f) for the 
decays \bzKz\ (a, b), \bppiz\ (c, d), and \bzpiz\ (e, f).
The solid lines represent the results of the fits, and the dashed and
dot-dashed lines the signal and background contributions respectively.
}
\end{figure}

The functions ${\cal P}_j$ are constructed as linear combinations of
Gaussian and 
polynomial functions, or in the case of \mes\ for \qqbar
background, the threshold function
$x\sqrt{1-x^2}\exp{\left[-\xi(1-x^2)\right]}$, with argument
$x\equiv2\mes/\sqrt{s}$ and shape parameter $\xi$.  These functions are
discussed in more detail in \cite{PRD04}, and are illustrated in
Figs.~\ref{fig:proj_b1Ks}\ and \ref{fig:proj_all}.

We allow the parameters most important for the determination of the
combinatorial background PDFs to vary in the fit, along with the
yields for all components, and the signal and
\qqbar\ background asymmetries.  Specifically, the free
background parameters are: $\xi$ for \mes, linear and quadratic
coefficients for \DE, and the mean, width, width difference, and
polynomial fraction parameters for \xf.

\begin{table*}[btp]
\caption{
Number of events $N$ in the sample, fitted signal yield $Y_S$, and 
measured bias (to be subtracted from $Y_S$) in events (ev.), detection
efficiency times secondary decay branching fractions $\epsilon$, 
significance~$\cal S$ (with systematic uncertainties included), and
branching fraction and charge asymmetry with
statistical and systematic error.
}
\label{tab:results}
\newcommand{\mn}{\ensuremath{\phantom{-}}}
\newcommand{\on}{\ensuremath{\phantom{1}}}
\newcommand{\eff}{$\epsilon$ (\%)}
\newcommand{\pbf}{$\prod\calB_i$ (\%)}
\newcommand{\signf}{$\cal S$ ($\sigma$)}
\begin{tabular}{lcr@{}lr@{}lccll}
\dbline
Mode	      	& $N$ (ev.)
			&\multicolumn{2}{c}{~$Y_S$ (ev.)~}
						&\multicolumn{2}{c}{~Bias (ev.)~}
									&\eff	
											&~~\signf
 													&\multicolumn{1}{c}{\calB\ $(10^{-6})$}	
															&\multicolumn{1}{c}{\acp}	\\
\tbline
\fbpKz		& 9841	&   $164$&$^{+27}_{-25}$	&    $15$&$\pm7$	& 3.4	& $\mn\sbpKz$	& $\on\rbpKz$	& \abpKz	\\
\fbzKz		& 5420	&    $58$&$^{+19}_{-17}$	&     $5$&$\pm3$
& 2.2	& $\mn\sbzKz$	& $\on\rbzKz\ (<\ulbzKz)$	
& 	\\
\fbppiz		& 28787	&    $71$&$^{+35}_{-32}$	&     $8$&$\pm4$	& 7.7	& $\mn\sbppiz$	& $\on\rbppiz\ (<\ulbppiz)$	& 	\\
\fbzpiz		& 10554	&     $6$&$^{+19}_{-16}$	&    $-2$&$\pm2$	& 4.8	& $\mn\sbzpiz$	& $\on\rbzpiz\ (<\ulbzpiz)$	& 	\\
\dbline
\end{tabular}
\end{table*}

We validate the fitting procedure by applying it to ensembles of
simulated experiments with the \qqbar\ component drawn from the PDF,
into which we have embedded known numbers of signal and \BB\ background
events randomly extracted from the fully simulated MC samples.  By
tuning the number of embedded events until the fit reproduces the yields
found in the data, we determine the biases that are reported, along with
the signal yields, in Table~\ref{tab:results}.

In Figs.\ \ref{fig:proj_b1Ks}\ and \ref{fig:proj_all}\ we show the
projections of the PDF and data for each fit.  The data plotted are
subsamples enriched in signal with the requirement of a minimum value of
the ratio
of signal to total likelihood (computed without the plotted variable)
that retains (30--50)\%  of the signal, depending on the mode. 

We compute the branching fraction by subtracting the fit bias from the
measured yield, and dividing the result by the number of produced \BB
pairs and by the efficiency times
${\cal B}(\omega\ra\pip\pim\piz)=89.1\pm0.7\%$ (and for the modes with
\KS, ${\cal B}(\Kz\ra\KS\ra\pip\pim)=\half(69.20\pm0.05)\%$)
\cite{PDG2006}.
The efficiency is obtained from the MC signal model. 
We assume that the branching fractions of the \UfourS\
to \BpBm\ and \BzBzb\ are each equal to $0.5$,
consistent with measurements \cite{PDG2006}.
The results are given in Table~\ref{tab:results},  
along with the significance, computed as the square root of the difference
between the value of $-2\ln{\cal L}$ (with additive systematic
uncertainties included) for zero signal and the value at its minimum.

Systematic uncertainties on the branching fractions arise from the PDFs,
\BB\ backgrounds, fit bias, and efficiency.  PDF uncertainties not
already accounted for by free parameters in the fit are estimated from
the consistency of fits to MC and data in control modes.   Varying the
signal-PDF parameters within these errors, we estimate yield
uncertainties of (1.6--6.4)\%, depending on the mode.  We estimate the
uncertainty of the MC model of misreconstructed signal by performing
alternate fits with a signal PDF determined from true signal events
only; we find differences of 1-4 events between these and the nominal fits.
The uncertainty  
from fit bias (Table \ref{tab:results}) includes its statistical
uncertainty from the simulated experiments, and half of the correction
itself, added in quadrature.  For the \BB\ backgrounds we vary the
fixed fit component by 100\% and include in quadrature a term derived from 
MC studies of the inclusion of a $b\to c$ component with the dominant
\qqbar\ background.
Uncertainties in our knowledge of the efficiency
include $0.5\%\times N_t$ and $1.5\%\times N_\gamma$, where
$N_t$ and $N_\gamma$ are the numbers of tracks and photons, respectively,
in the \B\ candidate.  
The uncertainties in the efficiency from the event selection are
below 0.5\%.

We study asymmetries from the track reconstruction (found to be negligible),
and from imperfect modeling of the interactions with material in the
detector, by measuring the asymmetries in the \qqbar\ background in
the data and control samples mentioned previously, in comparison with
MC \cite{KpiDCP}. We assign a systematic error for \acp\ equal to 0.01.

With the assumption that ${\cal
B}(\bone\ra\omega\pi) = 1$, we obtain
for the branching fractions (in units of $10^{-6}$):
\begin{eqnarray*}
\BbpKz  &=&  \rbpKz  \\
\BbzKz &=& \rbzKz~(<\ulbzKz) \\
\Bbppiz  &=&  \rbppiz~(<\ulbppiz)   \\
\Bbzpiz &=& \rbzpiz~(<\ulbzpiz). 
\end{eqnarray*}
The first error quoted is statistical and the second systematic.  We
find no evidence for the modes with \piz; the evidence for \BbzKz\ has a
significance of 3.4 standard deviations.  For these modes we quote also
90\% confidence level upper limits, given in parentheses.  We
observe the decay \BbpKz, and measure the charge asymmetry
$$
\begin{array}{ccc}
\acp(\bpKz) &=& \abpKz. 
\end{array}
$$ 

The QCD factorization estimates \cite{ChengYang}\ for the branching
fractions and charge asymmetry (0.014) agree with these measurements within
experimental and theoretical errors.  We find no evidence
for direct \CP\ violation in \BbpKz. 

\par We are grateful for the excellent luminosity and machine conditions
provided by our \pep2\ colleagues, 
and for the substantial dedicated effort from
the computing organizations that support \babar.
The collaborating institutions wish to thank 
SLAC for its support and kind hospitality. 
This work is supported by
DOE
and NSF (USA),
NSERC (Canada),
CEA and
CNRS-IN2P3
(France),
BMBF and DFG
(Germany),
INFN (Italy),
FOM (The Netherlands),
NFR (Norway),
MES (Russia),
MEC (Spain), and
STFC (United Kingdom). 
Individuals have received support from the
Marie Curie EIF (European Union) and
the A.~P.~Sloan Foundation.

%

\renewcommand{\baselinestretch}{1}


\begin{thebibliography}{99}
\bibitem{conjugate}
Charge-conjugate reactions are implied unless noted.
\end{thebibliography}
\end{document}